# Helio 2024 Science White Paper

# Ground-based Synoptic Studies of the Sun

Primary author:

V. Martinez Pillet (vmpillet@nso.edu), National Solar Observatory, 3665 Discovery Drive. Boulder, CO 80303 (orcid 0000-0001-7764-6895)

Co-authors:

A. Pevtsov (0000-0003-0489-0920), S. Gosain (0000-0002-5504-6773), National Solar Observatory, USA.

H. Gilbert (0000-0002-9985-7260), S. Gibson (0000-0001-9831-2640), A. G. de Wijn (0000-0002-5084-4661), J. Burkepile (0000-0002-9959-6048), High Altitude Solar Observatory, USA.

A. Asai (0000-0002-5279-686X), Kyoto University, Japan.

H. M. Bain (0000-0003-2595-3185), CIRES/CU-Boulder &N NOAA/SWPC, USA.

C. J. Henney (0000-0002-6038-6369), Air Force Research Laboratory, USA.

Y. Katsukawa (0000-0002-5054-8782), NAOJ, Japan.

H. Lin (0000-0002-4504-3165), Univ. Hawaii, USA.

W. Manchester (0000-0003-0472-9408), Univ. Michigan, USA.

J. McAteer (0000-0003-1493-101X), New Mexico State University, USA.

K. Muglach (0000-0002-5547-9683), Goddard Space Flight Center, USA.

M. Rast (0000-0002-9232-9078), LASP/CU Boulder, USA

M. Roth (0000-0002-1430-7172), Thüringer Landessternwarte, Germany.

J. Zhang (0000-0003-0951-2486), George Mason University, USA

**Primary Category**: Basic Research, Solar Physics
**Secondary Category**: Space Weather

**Synopsis**
*Ground-based synoptic solar observations provide critical contextual data used to model the large-scale state of the heliosphere. The next decade will see a combination of ground-based telescopes and space missions that will study our Sun's atmosphere microscopic processes with unprecedented detail. This white paper describes contextual observations from a ground based network needed to fully exploit this new knowledge of the underlying physics that leads to the magnetic linkages between the heliosphere and the Sun. This combination of a better understanding of small-scale processes and the appropriate global context will enable a physics-based approach to Space Weather comparable to Terrestrial Weather forecasting.*

**1.  A multi-messenger approach to understanding a connected heliosphere**
Stars and planets interact through a diverse suite of messengers: radiation in the form of photons, electromagnetic fields, and charged particles. NASA's Heliophysics System Observatory, the NSF's Daniel K. Inouye Solar Telescope (DKIST), and other ground-based assets use these multi-messenger signals to study the Sun-Earth interactions. Projects like DKIST target the **microphysics** that create the magnetic connectivity between the Sun and the heliosphere. To fully exploit these detailed observations and to drive heliophysics discovery over the next decade, the **large-scale context** provided by synoptic observations is of critical relevance. Existing ground-based synoptic programs are aging rapidly and are used in ways that differ from their original intend. Most prominently, GONG (Hill, 2018) was designed for Helioseismology but is most demanded today as a provider of the magnetic boundary conditions for solar wind models. Moreover, a wealth of theoretical knowledge about the connectivity between the Sun and the heliosphere has emerged recently (e.g., Chen 2017). This knowledge did not exist at the time of the design of existing synoptic networks like GONG, and they do not help predict key Space Weather drivers such as the magnetic field configuration of coronal mass ejections (CMEs), which determines the coupling with the Earth's magnetic field (Kilupa, 2017).

***The next decade will see fundamental progress in our ability to describe the magnetic connections existing throughout the heliosphere. Most importantly, the community will develop the tools to routinely propagate magnetized CMEs using appropriate boundary conditions on the Sun and enabling the prediction of their Bz, the southward magnetic field component in the CME arriving at Earth.***

**2.  Boundary data to propagate magnetic fields into the heliosphere**
Global 3D semi-empirical models simulating the heliosphere such as the WSA (Wang and Sheeley, 1990; Arge et al., 2003) use synoptic magnetic field maps to model a time-dependent solar wind. Much of the space weather activity can only be interpreted in light of such global 3D simulations (Bain et al., 2016). Investigations are underway to use heliospheric solar wind models to drive simulations of the geomagnetic impacts of space weather at the Earth, thus coupling space weather phenomena across the entire Sun-Earth system. The models are further utilized by injecting, on top of the solar wind, a solar wind disturbance (representing a CME) at the inner boundary and propagating it throughout the solar system.

CMEs are prevalent at solar maximum. During the minimum of the cycle, the open magnetic flux at the coronal holes near the Sun's poles establishes a quasi-dipolar magnetic configuration in the heliosphere. In this configuration, the Earth stays most of the time magnetically connected to the polar caps (Luhmann et al., 2009). Unfortunately, these coronal regions are the hardest to observe from the Earth. However, detailed observations from Hinode of the polar regions in the lower chromosphere show network-like concentrations displaying an apparent neutral line parallel to the limb reflecting the field line opening into the corona (red oval in Figure 1, bottom images). This expansion of the flux tubes —or expansion factor— relates to the speed of the plasma outflowing from the Sun (Figure 2, Sheeley, 2017).

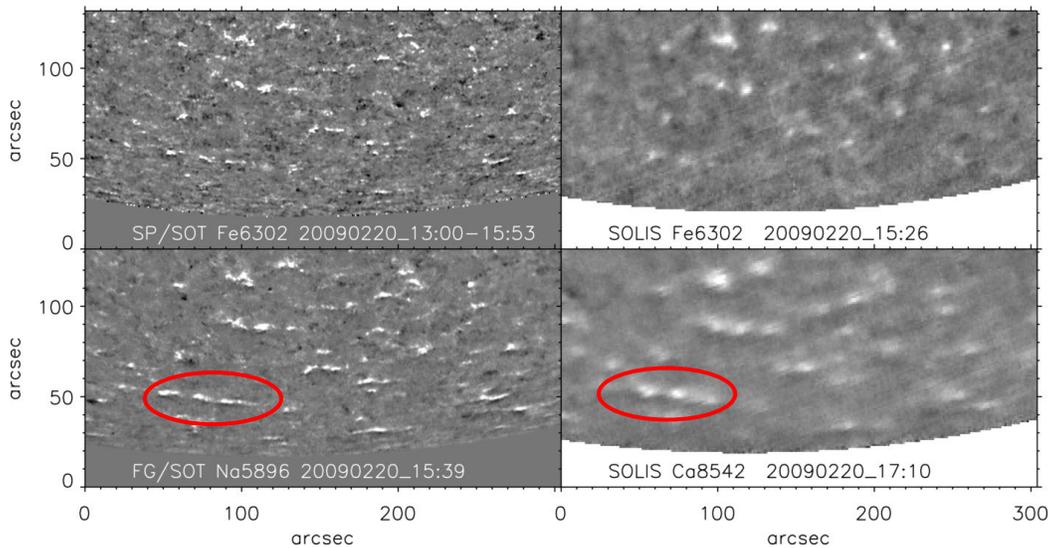

Figure 1: Nearly simultaneous south pole BLOS observations with the pole tipped toward Earth. Left: Hinode observations. Right: SOLIS observations. The ellipses in the bottom images show facular regions with apparent neutral lines created by the field line expansion (Jin et al., 2013).

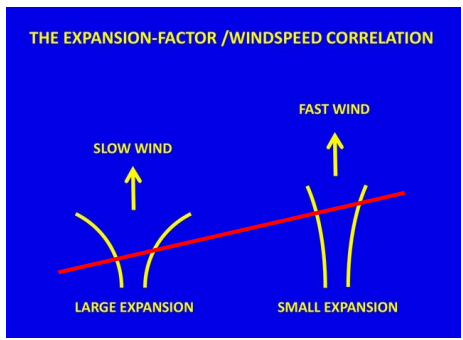

Figure 2. A cartoon illustrating that fast and slow winds come from magnetic flux tubes with different flux tube expansion. The red line demonstrates that the projection of expanding field lines in opposite directions will create an apparent neutral line effect (Sheeley, 2017).

As shown on the bottom right panel of Fig. 1, SOLIS chromospheric line-of-sight data also detects this field line expansion but with a degraded spatial resolution. Note that the apparent neutral lines are hardly observable in the corresponding photospheric magnetograms (Figure 1, top images). Unfortunately, the synoptic magnetograms that feed current heliospheric solar wind models (GONG, SDO/HMI) observe only the photosphere and do not contain these details. SOLIS chromospheric magnetograms have not been used yet by the heliospheric modeling community. However, the potential benefits for solar wind modeling using data that directly observes the expansion factors, instead of estimating them from a current-free extension of the photospheric field, are apparent.

Projection effects make it difficult to observe the solar poles from the ecliptic plane, but targeted, carefully designed observations from the ground can significantly improve our boundary data from these regions. The polar coronal holes occupy an extended area at latitudes higher than 60º-70º. Outside of the times when the polar fields undergo polarity reversals, the magnetic configuration is relatively simple with prevailing vertical field lines. From the Earth's line of sight, this configuration corresponds to **transverse** magnetic field observations. The Zeeman effect makes the transverse signals smaller than longitudinal ones, often by as much as one order of magnitude. Improving our sensitivity to transverse magnetic fields and beating the noise in polar cap observations from the ground demands detecting more polarized photons than what we currently collect.

***Obtaining quantitatively relevant polar maps from the Earth line-of-sight demands higher spatial resolution and increased magnetic sensitivity to transverse fields using new instrumental approaches for synoptic observations.***

## 3. Boundary data and the propagation of solar eruptive structures

While the background solar wind is calculated from photospheric magnetic field synoptic maps, the CME modeling component uses a 'cone model' (Odstrcil & Pizzo, 1999) fed by parameters derived from coronagraph observations. These parameters contain no information of the CME magnetic field configuration resulting in purely hydrodynamic modeling of the cloud. This approach estimates arrival time, CME speeds, and densities. However, it contains **limited or no** information about the magnetic configuration of the cloud and its potential for interaction with the heliosphere and the magnetic field of the planets it encounters.

Most of the widely used 3D global heliospheric models that propagate CMEs do not propagate magnetized CMEs and cannot predict the value of Bz. However, in recent years capable, physics-based models are appearing (Shiota & Kataoka, 2016; Jin et al., 2017; Torok et al., 2018; Verbeke et al., 2019; Sarkar et al., 2020) and contain the ability to propagate a magnetized CME from the solar surface to the heliosphere (Figure 3). Given the complexity of the required modeling, these studies consider one-off CME examples. ***However, quantifiable forecasting of Bz requires routine propagation of magnetized CMEs from the solar source into the heliosphere using adequate boundary data.*** Earth-directed CMEs are unpredictable and require constant monitoring of all potential source regions on the visible surface of the Sun, demanding a network that avoids night gaps. As inferred from the studies mentioned above, the necessary boundary data is a rich combination of vector magnetic fields in the photosphere and chromosphere that permits the characterization of the overall configuration and the helicity of the source regions; Doppler data of the dense filaments at the start of the ejection (Wang et al., 2022); and a description of the coronal environment where the eruption occurs and that potentially impacts its evolution (Corchado-Albelo et al., 2021).

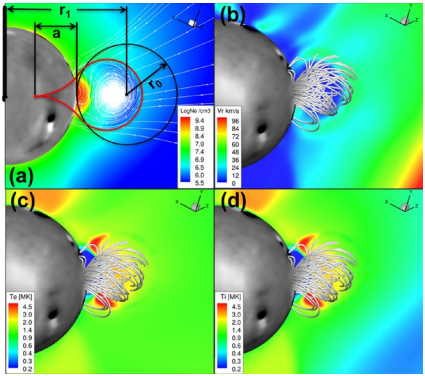

*Figure 3. Parametrized model for the propagation of a magnetized CME (Jin et al., 2017). Initial flux rope configuration embedded into the global solar wind solution of CR 2107. (a) Flux rope in 2D with magnetic field lines and plasma density. The black and red contours represent unstretched and stretched flux ropes, respectively. (b)–(d): Flux rope in 3D with the radial velocity, electron temperature, and proton temperature, respectively.*

Chromospheric vector fields are a primary candidate for improving boundary data (Georgoulis et al., 2018). Mapping the field configuration at multiple heights will produce data for constraining the magnetic topology of solar filaments associated with the core of the CME and will generally serve as improved initial boundary conditions for CME modeling. We expect interactions with the background corona during and after the eruption to lead to CME deflections, etc. However, despite these complexities, observed persistent correlations demonstrate that CMEs maintain a non-negligible memory of their solar source region (Yurchyshyn et al., 2001; Marubashi et al., 2015; Aparna and Martens, 2020).

***These correlations indicate that data-driven propagation of magnetized CMEs is a real possibility but needs consistent boundary data that is not currently available***.

## 4. Coronal data to inform and constrain CME models.

The nature of the CME source region, including the presence of stored magnetic energy in the form of a magnetic flux rope, can have signatures in coronal observations such as prominences and their cavities (Gibson, 2018). Ground-based synoptic coronagraphic observations are crucial to analyzing the evolution leading up to the eruption, with important information from the morphology, plasma properties, and polarimetry of CME source regions. During an eruption, such observations yield global evidence for the effects described above, which may alter the pre-eruption magnetic structure and the Bz at Earth.

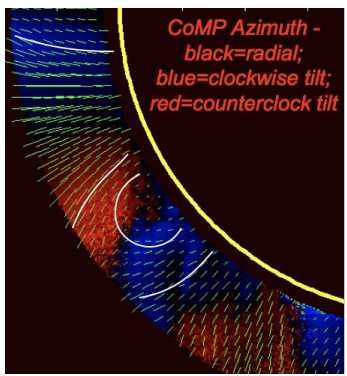

*Figure 4: CoMP observations of linear polarization azimuth in FeXIII. In the region of diverging magnetic field (top left) the polarization vectors are parallel to the magnetic field and smoothly transition from blue (tilted clockwise from radial) to red (tilted counterclockwise from radial). In the region of converging magnetic field (middle), the polarization vectors follow the magnetic field direction until they cross the van Vleck angle (near the center of the loops), at which point they abruptly rotate radial. The white lines can be thought of as tracing the streamer morphologies observed by K-Cor.*

Coronagraphic observations provide vital measurements that can be incorporated into first principles models for use in ensemble forecasting of space weather conditions. Several groups are developing research pipelines to facilitate incorporation into heliospheric models with ensemble down-selection using observed coronal morphology, and, more generally, the validation of magnetic extrapolations and MHD models. Jones et al. (2020) developed a technique for comparing structures observed in K-Cor (Mauna Loa) white-light pB coronagraph images to WSA magnetic models for use in boundary data conditioning and ADAPT (Hickmann et al., 2015) parameter determination. CoMP (Tomczyk et al., 2008) azimuth also diagnoses magnetic field morphology, which is important for constraining solar wind models, as shown in Figure 4. In addition, CoMP linear polarization observations have been used to diagnose coronal magnetic field direction and topology (Bak-Steslicka et al., 2013; Rachmeler et al., 2014; Gibson et al., 2017, Corchado-Albelo et al., 2021). CoMP Doppler measurements of Alfvenic waves have also been used to quantify plane-of-sky magnetic field strength (Yang et al., 2020).

***Coronal observations provide ground truth validation and constraints on models of the solar wind and CMEs.***

## 5. Helioseismology as a window to the sources of solar magnetism

The origin of the solar magnetism described above lies in the interior, underneath the visible photosphere. This region is where large-scale motions of the plasma generate the field via, presumably, a dynamo mechanism. Helioseismology—analysis of properties of acoustic waves traveling in the solar interior—is the best probe of these motions.

While helioseismology has succeeded in determining the large-scale flows averaged over the entire Sun, it has not successfully determined the structure and dynamics below magnetized regions. For example, wave-speed perturbations beneath an active region obtained from time-distance inversions differ from ring-diagram inversions (Gizon et al., 2009) and disagree with the results from semi-empirical models (e.g., Moradi et al., 2010). These disagreements indicate inadequate treatment of wave behavior near strong and inclined magnetic fields in the helioseismic analysis. The missing ingredient is currently thought to be the transformation of the acoustic waves used for helioseismology into additional MHD oscillation modes (Alfvén waves, fast magnetic, slow magnetic, etc.) at the height in the atmosphere where $\beta$, the ratio of thermodynamic gas pressure to magnetic pressure, equals 1. This level marks the boundary between atmospheric regions where the magnetic field dominates the plasma motions ($\beta < 1$) and regions where the gas pressure is the primary driver ($\beta > 1$).

There has been substantial theoretical and observational work on this problem in the last decade (e.g., Cally, Moradi & Rajaguru 2016). These studies have sought a method of correcting the helioseismic analyses in active regions. While there may well be further improvements, the current approach is to forward model the effect using a sunspot background model with embedded acoustic sources (Moradi et al., 2015). This forward modeling produces estimates of the perturbation to the travel time of waves passing through the region as a function of field strength, inclination, azimuthal angle, height, input wave frequency, and travel distance. These perturbed travel times can then be applied to the relevant time-distance helioseismic analyses. With additional theoretical work plus routine multi-wavelength, multi-height helioseismic observations, we may reach the long-standing promise of **detecting active regions on the Sun before they emerge** to the surface.

Helioseismic holography is a technique used to infer active regions on the **far-side** surface of the Sun (Lindsey & Braun, 2017). This approach exploits acoustic travel-time reductions in magnetized areas that result in a phase shift of the waves. These phase changes make large active regions readily apparent in reconstructed seismic travel-time images (Figure 5). Since the sensitivity in these maps depends on accurate and precise measurements of the phase shift between acoustic waves in the solar atmosphere, an improved understanding of phase shift from multi-height observations is needed to reduce the noise in far-side maps, thus enhancing the detectability of weaker active regions and improving the modeling of the global heliosphere.

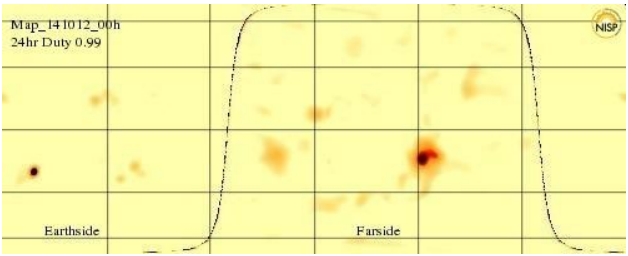

*Figure 5: A map of the far-side of the Sun produced from helioseismic observations from GONG on Oct 12, 2014. The sizeable prominent feature on the far-side is Active Region AR12192.*

***Observations required to address the detectability of these phase shifts include multi-wavelength, multi-height measurements of the velocity and vector magnetic field from the photosphere to the chromosphere with cadences of minutes as needed by helioseismology.***

**6. Global images as the basis for total and spectral irradiance modeling**

The most fundamental connection between the sun and the Earth occurs via its radiative output. While global climate change is dominated by anthropogenic contributions, solar irradiance variation induces climate modulation. For example, the Sun played a role in the slower rate of global surface temperature rise during the first decade of this century (IPCC Climate Change 2021: The Physical Science Basis). The mechanisms underlying these contributions remain uncertain. In particular, cycle variations of the solar spectral irradiance are still debated, with widely varying measured and modeled contributions as a function of wavelength (e.g., Ermolli et al 2013). Understanding these is critical to understanding the details of climate response, particularly at regional scale.

The solar radiative output varies with magnetic field and with the position of magnetic field structures on the solar disk. Sunspots and pores have negative contrast against the background disk, while smaller scale magnetic fields generally contribute positively. Changes in the fractional areas covered by structures of different intensities modulate the solar irradiance. The magnitude and even the sign of the contribution of any given magnetic structure, however, depends also on the wavelength being observed and the structure's location on the solar disk. To understand solar irradiance variations three activities needed: 1) accurate and precise observations of the evolution of active regions on the Sun through global solar imaging. 2) models and high-resolution observations of solar magnetic structures that explain their radiant emission and its variance. 3) the measurement of disk integrated solar spectral irradiance against which models can be tested.

Precision photometric imaging of the Sun (measuring the intensity of a pixel on the solar disk relative to the background) is required to carefully understand the spectral irradiance contributions of the magnetic field. To date such data, taken by the Precision Solar Photometric Telescope (PSPT; Rast et al. 2008) network, is available for the period 1998 to 2015 in three wavelength bands (red and blue continua and Ca II K). By careful determination of the detector gain and the quiet-sun center-to-limb variation, the PSPT achieved high precision (0.1%) relative photometry of the disk. Extension of this synoptic measurement is essential to understand decades long variation in solar spectral output and their impact on terrestrial climate.

## 7. Required solar synoptic observations over the next decade

Key measurement capabilities for future ground-based solar synoptic network facilities involving the US and the broader international communities include the following:

- Measure the **boundary data** as a function of height that propagates the magnetic connectivity into the heliosphere, with improved sensitivity to the **polar regions**;
- Map coronal magnetic field topology and plasma properties to **improve solar wind** models;
- Map the **3-D magnetic topology** of solar erupting structures from the chromosphere to the corona and better anticipate the severity of space-weather events;
- Provide plasma diagnostics of ambient and eruptive coronal structures to study energy build-up and release in the corona and **improve CME** modeling;
- Monitor processes in the **solar interior** and the **far side** that impact heliospheric conditions;
- Provide global **context** for high-resolution solar observations as well as for *in situ* measurements; and
- Extend in time the data base of solar disk images with quantifiable high photometric accuracy.

The use of a six-site network guarantees a duty cycle of 90% as demonstrated by Jain et al. (2021).

**The ongoing design efforts for the Next Generation GONG (ngGONG) network incorporates uniquely these capabilities by hosting a suite of telescopes able to observe the photosphere, chromosphere, and the plane of the sky corona at six longitudes from the ground. ngGONG is further described in the WP Pevtsov et al. (2022).**